\documentclass[prb,showpacs,twocolumn,amsmath,amssymb,groupadress,superscriptaddress]{revtex4}
\usepackage{graphicx}% Include figure files
\usepackage{dcolumn}% Align table columns on decimal point
\usepackage{bm}% bold math

\begin{document}

\preprint{APS/PRB -Khalyavin}

\title{Spin-ordering and magnetoelastic coupling in the extended Kagome system YBaCo$_4$O$_7$}

\author{D.D. Khalyavin}
\affiliation{ISIS facility, Rutherford Appleton Laboratory-CCLRC,
Chilton, Didcot, Oxfordshire, OX11 0QX, UK. }
\author{P. Manuel}
\affiliation{ISIS facility, Rutherford Appleton Laboratory-CCLRC,
Chilton, Didcot, Oxfordshire, OX11 0QX, UK. }
\author{B. Ouladdiaf}
\affiliation{Institute Laue-Langevin, 6 Rue Jules Horowitz, BP 156, 38042 Grenoble Cedex 9, France }
\author{A. Huq}
\affiliation{Materials Science Division, Argonne National Laboratory, Argonne, Illinois 60439, USA. }
\author{H. Zheng}
\affiliation{Materials Science Division, Argonne National Laboratory, Argonne, Illinois 60439, USA. }
\author{J.F. Mitchell}
\affiliation{Materials Science Division, Argonne National Laboratory, Argonne, Illinois 60439, USA. }
\author{L.C. Chapon}
\affiliation{ISIS facility, Rutherford Appleton Laboratory-CCLRC,
Chilton, Didcot, Oxfordshire, OX11 0QX, UK. }
\date{\today}% It is always \today, today,

\begin{abstract}
Low temperature magnetic and structural behavior of the extended Kagome system YBaCo$_4$O$_7$ has been studied by single crystal neutron diffraction and high-resolution powder X-ray diffraction. Long-range magnetic ordering associated with a structural transition from orthorhombic $Pbn$2$_1$ to monoclinic $P$2$_1$ symmetry has been found at T$_1$ $\sim$ 100 K. The interplay between the structural and magnetic degrees of freedom testifies that the degeneracy of the magnetic ground state, present in the orthorhombic phase, is lifted through a strong magnetoelastic coupling, as observed in other frustrated systems. At T$_2$ $\sim$ 60 K, an additional magnetic transition is observed, though iso-symmetric. Models for the magnetic structures below T$_1$ and T$_2$ are presented, based on refinements using a large number of independent reflections. The results obtained are compared with previous single crystal and powder diffraction studies on this and related compositions.
\end{abstract}

\pacs{75.25.-j, 75.85.+t, 61.05.F-}% PACS, the Physics and Astronomy
\maketitle

\section{Introduction}
\indent Magnetic behavior of systems with competitive exchange interactions has been the focus of attention over the last several decades. This competition can arise due to comparable strength between nearest and beyond nearest neighbor interactions or due to specific symmetry of the exchange topology.\cite{ISI:000235221400014} The latter, usually referred to as geometrical frustration, can be found in many materials where magnetic ions form a network of triangular-based structural units, as for example in two-dimensional Kagome and three-dimensional pyrochlore lattices.\cite{ISI} The \emph{hallmark} of frustration is a large magnetic ground state degeneracy and consequently the suppression of long-range magnetic ordering. This degeneracy is possibly lifted by structural distortions \cite{ISI:000173849300056,ISI:000177911700058,ISI:000230276600010,ISI:000257288900082} either associated to non magnetic instability (bonding requirements, orbital/charge order) or as a consequence of strong magnetoelastic coupling. In the resulting distorted phases, complex non-collinear or partially disordered spin configurations can be observed.\cite{ISI:000223017300001,ISI:000177911700058} Recently, this kind of behavior has been reported for $R$BaCo$_4$O$_7$ ($R$-rare earth or Y) cobaltites.\cite{ISI:000236501300023,ISI:000242409000015} These compositions belong to a new class of frustrated systems with trigonal symmetry ($P31c$), formed by an alternate stacking of Kagome and triangular layers along the $c$-axis.\cite{ISI:000177955700007} The unique exchange topology is formed by a network of corner-sharing trigonal bipyramids and triangular clusters.\cite{ISI:000268088300060} A structural phase transition from trigonal ($P31c$) to orthorhombic symmetry ($Pbn2_1$) has been observed for most compositions, driven by chemical bonding conditions of Ba$^{2+}$ and $R^{3+}$ cations. The temperature of the $P31c$ $\rightarrow$  $Pbn$2$_1$ phase transition was found to vary from 160 K for $R$=Lu up to 355 K for $R$=Ho.\cite{ISI:000236501300023,ISI:000242409000015,ISI:000241855500005,ISI:000236596100022,ISI:000276929900003} Short range magnetic correlations, evidenced by diffuse neutron scattering, are drastically enhanced in the orthorhombic phase,\cite{ISI:000268088300060} but there is still a controversy about the nature of the low-temperature magnetic state. Soda et al.\cite{ISI:000237969600046} reported only short-range correlations in YBaCo$_4$O$_{7+\delta}$ single crystal and discuss models where the Kagome and triangular layers are fully decoupled, forming spin configurations close to 120$^\circ$. A similar picture of finite magnetic correlations and decoupled Kagome and triangular spins has been adopted to interpret the neutron diffraction measurements on LuBaCo$_4$O$_{7+\delta}$ single crystal.\cite{ISI:000249279100024} In contrary, long range magnetic ordering with an intricate magnetic structure has been found for a polycrystalline sample of YBaCo$_4$O$_7$ below 110 K.\cite{ISI:000242409000015} The discrepancy with the single crystal data is possibly due to the oxygen stoichiometry in the crystals studied by Soda et al.\cite{ISI:000237969600046,ISI:000249279100024}, an assumption based on the observation that hyper-stoichiometric Y-based and related compositions\cite{ISI:000236501300023} which do not undergo the structural transition to the orthorhombic phase, remain short-range ordered. On the other hand, the $^{57}$Fe Mossbauer spectra of both YBaCo$_{3.94}$Fe$_{0.04}$O$_{7.02}$ and YBaCo$_{3.94}$Fe$_{0.04}$O$_{7.80}$ polycrystalline samples collected at 4K shows well defined magnetic sextets consisting of two and three components respectively,\cite{ISI:000264302000034} a result fully compatible with the presence of long-range magnetic order in both these compositions.\\
\indent With the aim to resolve the controversy about the nature of the magnetic state in YBaCo$_4$O$_7$, we undertook additional neutron diffraction experiments on single crystal with an oxygen content precisely controlled. Our results are consistent with the scenario of long-range magnetic order, evidenced by the presence of sharp magnetic Bragg peaks. A model for the magnetic structure is proposed from refinements based on a large number of independent reflections. In addition, high resolution powder X-ray diffraction data clearly indicates that the symmetry of the magnetically ordered state is lowered to monoclinic $P2_1$. The latter observation testifies of the strong spin-lattice coupling in the system and its inherent link to the magnetic frustration not fully lifted by the orthorhombic $Pbn2_1$ symmetry. 
\section{Experimental part}
\begin{figure}[t]
\includegraphics[scale=0.95]{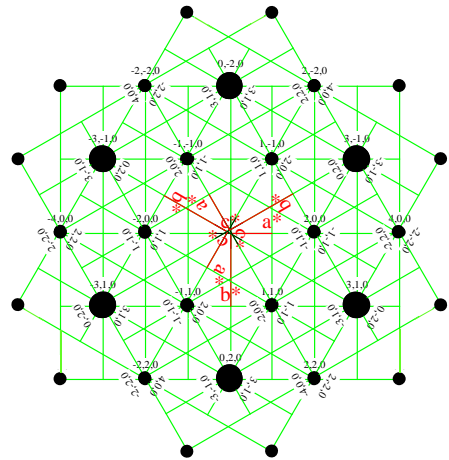}
\caption{(Color online) A part of the ($hk0$) reciprocal plane demonstrating superposition of the three types of 120$^\circ$ ferroelastic domains, expected in the case of the orthorhombic symmetry.}
\label{fig:dom}
\end{figure}

\indent The single crystal of YBaCo$_4$O$_{7.0}$ was grown by a floating zone technique in an optical image furnace. A densified rod of the nominal composition was melted in a 20\% O$_2$/Ar atmosphere, and the crystal was grown at 1 mm/h. Because of the high affinity for YBaCo$_4$O$_7$ to pick up oxygen on cooling, an afterheater was used to keep the growing crystal above 900$^\circ$C. After growth was complete, the atmosphere was changed to 99.995\% Ar for 24 h and the crystal cooled down to room temperature over six hours. Oxygen content was verified by thermogravimetric analysis of a crushed sample of the crystal. The ceramic samples used for the single crystal growth and X-ray powder diffraction experiments were synthesized from high-purity Y$_2$O$_3$, BaCO$_3$, and Co$_3$O$_4$ reagents by repeated firing in air at 1150$^\circ$C. The materials for the X-ray measurements were then treated in nitrogen at 1150$^\circ$C resulting in the oxygen content O$_{6.95(5)}$ as determined by thermogravimetry.\\
\indent The neutron diffraction studies were performed on the PRISMA time-of-flight spectrometer at the ISIS facility, Rutherford Appleton Laboratory (U.K.) and on the four-circle diffractometer D10 at the ILL (Grenoble, France). For the experiment at ISIS, the single crystal of cylindrical shape (5 mm diameter, 3 mm height) was mounted on an aluminium pin with the [001] vertical axis. For this geometry, scattering in the orthogonal (\textit{a$^*$b$^*$}) reciprocal plane was recorded by rotating the crystal around the [001] axis. Data have been corrected for absorption and normalized to the incoherent scattering of a vanadium standard. The experiment at ILL was done on the same crystal prealigned by using the OrientExpress facility. All data were collected with an incident neutron wavelength of $\lambda $=2.36 $\AA$ by using an 80 mm$^2$ two-dimensional microstrip detector. Peak integration was performed using the program RACER (ILL) in two steps. First, a library was built by fitting ellipsoidal shapes to a set of strong reflections ($I>3\sigma $), these shapes were used in a second pass to integrate all reflections. For each data set the list of integrated intensities obtained were corrected for Lorentz factors and normalized to the monitor count. To refine the magnetic structure (propagation vector {\bf k}=0), 432 independent reflections were collected at 5 K and 80 K. In addition, intensity of some selected peaks was measured as a function of temperature (5 K$<$T$<$125 K). The nuclear structure in the paramagnetic phase was refined based on 266 independent reflections measured at 150 K. Refinement of both the nuclear and magnetic structures was carried out using the FullProf program.\cite{ISI:A1993ME99200007} \\
\indent A good agreement between the observed and calculated structure factors (RF=5.95\%) in the paramagnetic phase was obtained in the orthorhombic symmetry $Pbn2_1$ ($Pna2_1$ in standard setting) with three types of 120$^\circ$ ferroelastic domains shown in Figure \ref{fig:dom}.
These domains are expected due to the symmetry relation between the orthorhombic $Pbn2_1$ and hexagonal $P6_3mc$ space groups. The latter is the common supergroup of $Pbn2_1$ and $P31c$ and is the parent symmetry which has to be used for classification of domains and displacive modes in the system.\cite{ISI:000271351500043} Matrix representation of the rotational parts of the three domain generators: \{E/000\}, \{C$_{3^+}$/000\} and \{C$_{3^-}$/000\} were used to introduce the twining law at the data processing stage. The low temperature magnetic phases at 5 K and 80 K were refined in monoclinic $P$2$_1$ space group with six types of domains as discussed in the next section.\\
\begin{figure}[b]
\includegraphics[scale=1.0]{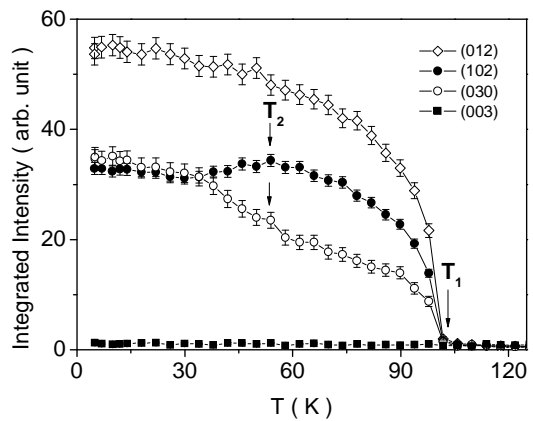}
\caption{Integrated intensities of some selected peacks as a function of temperature (error bars correspond to 3$\sigma$ interval). The indexation relates to the orthorhombic $Pbn$2$_1$ unit cell with $a\sim$6.30 $\AA$, $b\sim$10.95 $\AA$ and $c\sim$10.19 $\AA$.}
\label{fig:int}
\end{figure}
\indent High-resolution X-ray powder diffraction data were collected on beamline X16C at the National Synchrotron Light Source (USA). X-rays of wavelength 0.6911 $\AA$ were selected by a Si(111) double-crystal monochromator. Samples were mounted on a flat brass plate, and data were collected in the reflection geometry within the temperature interval 10 K$<$T$<$200 K.
\section{Results and discussion}
\begin{figure}[t]
\includegraphics[scale=1.0]{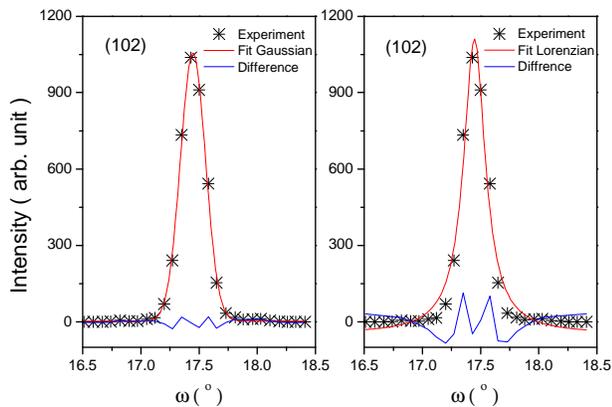}
\caption{(Color online) Example of $\omega$-scan profile for the (102) reflection measured at 5 K on the D10 diffractometer and fitted with a Gaussian (left) and Lorentzian (right) functions. The reflection, forbidden in the $Pbn2_1$ space group, has mainly a magnetic contribution (some small nuclear contribution can exist since it is allowed in $P112_1$).}
\label{fig:lor}
\end{figure}

\indent In agreement with the measurements on polycristalline materials,\cite{ISI:000242409000015} our single crystal neutron diffraction data clearly indicate two successive phase transitions in YBaCo$_4$O$_7$, which take place at T$_1\sim$100 K and T$_2\sim$60 K respectively (Fig.\ref{fig:int}); the former has been associated with long-range magnetic ordering (propagation vector {\bf k}=0) and the latter with a spin reorientation process.\cite{ISI:000242409000015} In the previous single crystal experiments,\cite{ISI:000237969600046,ISI:000249279100024} the magnetic reflections were found to have a Lorentzian shape and a width indicating short range spin correlations. The authors pointed out that the fundamental discrepancy with the powder diffraction data\cite{ISI:000242409000015} could be related to a less accurate estimation of the magnetic component in the powder experiments. In Figure \ref{fig:lor}, $\omega$-scan for the (102) reflection measured at 5 K is presented as an example of our single crystal data. This reflection is forbidden by the $n$-glide plane of the $Pbn2_1$ space group and therefore has mainly a   magnetic contribution. In addition, this particular reflection is not affected by the 120$^\circ$ domain structure which can introduce some broadening or asymmetry. The reflection has a perfect Gaussian shape and a width essentially resolution limited, indicating the long-range nature of magnetic ordering.\\
\begin{figure}[t]
\includegraphics[scale=1.0]{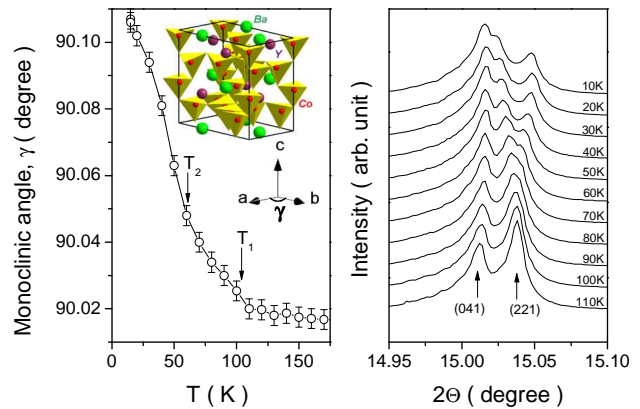}
\caption{(Color online) Monoclinic angle as a function of temperature (left panel). The inset shows the orthorhombic unit-cell and the angle $\gamma$ which deviates from 90$^\circ$ in the monoclinic phase. A part of the X-ray powder diffraction patterns collected at different temperatures and demonstrating the splitting of the (221) reflection below T$_1\sim$100 K (right panel).}
\label{fig:xrd}
\end{figure}
\subsection{Symmetry of the magnetically ordered state}
\indent The high-resolution powder X-ray diffraction reveals that magnetic ordering is accompanied by structural distortions that reduce the crystal symmetry down to monoclinic (Fig.\ref{fig:xrd}), producing a clear splitting of some of the nuclear reflections below T$_1\sim$100K. The systematic splitting of the ($hk0$) reflections and the absence of splitting for the ($h0l$),($0kl$) reflections indicate a unit-cell with $c$ as unique axis and $\gamma$ as monoclinic angle (Fig.\ref{fig:xrd} inset). For simplicity, this non-conventional setting will be used to describe the unit-cell rather than transforming the coordinates to the conventional cell. The temperature dependence of $\gamma$ is plotted in Fig. \ref{fig:xrd} (left). Above T$_1$, the small deviation of $\gamma$ from 90$^\circ$ is caused by anisotropic peak broadening which often precedes magnetostructural phase transitions.\cite{ISI:000251674300063} In fact, above T$_1$, refinements with monoclinic symmetry do not improve what is obtained with orthorhombic symmetry, indicating that the structural transition coincides with the magnetic one.
\indent The change of unit-cell metric at the magnetic transition is associated with the $e_{xy}$ component of the strain tensor, which spontaneously appears below T$_1$. This observation provides crucial information about the magnetic symmetry because a magnetoelastic coupling invariant, linear in respect of $e_{xy}$, must be present in the Landau expansion of the free energy. Since the bilinear coupling between $e_{xy}$ and the magnetic order parameter, $\eta$, is forbidden by time-reversal, one must consider coupling terms that involve even powers of the magnetic order parameter(s). Inspection of the irreducible representations (irreps) of the $Pbn2_1$ space group shows that there are four one-dimensional irreps listed in Table \ref{table:irrep}. The strain component $e_{xy}$ transforms according to the $\Gamma_2$ representation. It is obvious that only the direct products $\Gamma_1\otimes\Gamma_2$ or $\Gamma_3\otimes\Gamma_4$ $\ni\Gamma_2$ and therefore only coupling invariants of the form $e_{xy}\eta_1\eta_2$ or $e_{xy}\eta_3\eta_4$ are possible. It is alternatively easy to determine by direct inspection of the irreps table that the only way to loose the mirror symmetries whilst preserving the two-fold axis along $c$, necessary to stabilize this monoclinic symmetry, is to mix either $\Gamma_1$ and $\Gamma_2$ or $\Gamma_3$ and $\Gamma_4$. In the former case, the magnetic symmetry is $P112_1$ and in the latter, $P112_1'$. The systematic absence of (0 0 $l=2n+1$) magnetic reflections (Fig.\ref{fig:int}), is uniquely consistent with a time-reversed $2_1$ axis, and thus we can conclude that below T$_1$, the magnetic ordering has the symmetry of the reducible $\Gamma_3\oplus\Gamma_4$ order parameter.

\begin{table}[t]
\caption{Irreducible representations of $Pbn2_1$ space group associated with wave vector {\bf k}=0.}% title of Table
\centering % used for centering table
\begin{tabular*}{0.48\textwidth}{@{\extracolsep{\fill}} c c c c c} % centered columns (4 columns)
\hline\hline %inserts double horizontal lines
Irrep & $\lbrace 1 | 0 0 0 \rbrace$ & $\lbrace m_{yz} | \frac{1}{2} \frac{1}{2} 0 \rbrace$ & $\lbrace m_{xz} | \frac{1}{2} \frac{1}{2} \frac{1}{2} \rbrace$ & $\lbrace 2_{z} | 0 0 \frac{1}{2} \rbrace$ \\ [0.5ex] % inserts table
%heading
\hline % inserts single horizontal line
$\Gamma_1$ & 1 & 1 & 1 & 1 \\ % inserting body of the table
$\Gamma_2$ & 1 & -1 & -1 & 1 \\
$\Gamma_3$ & 1 & 1 & -1 & -1 \\
$\Gamma_4$ & 1 & -1 & 1 & -1 \\ 
\hline
\hline  
\end{tabular*}
\label{table:irrep} 
\end{table}
\begin{table}[b]
\caption{Atomic components of basis vectors, $\varphi_{i\alpha=x,y,z}$ of $\Gamma_3$ and $\Gamma_4$ irreducible representations of $Pbn2_1$ space group entering three times ($i$=3) in the reducible magnetic representation for an atom occupying the 4$a$ Wyckoff position of the $Pbn2_1$ space group.}
\centering % used for centering table
\begin{tabular*}{0.48\textwidth}{@{\extracolsep{\fill}} c c rrr rrr rrr rrr} % centered columns (4 columns)
\hline\hline %inserts double horizontal lines
4$a$ & & \multicolumn{3}{c}{1\footnotemark[1]} &  \multicolumn{3}{c}{2}  &  \multicolumn{3}{c}{3}  &  \multicolumn{3}{c}{4}  \\ [0.5ex] % inserts table %heading
\hline % inserts single horizontal line
$i$ & & 1 & 2 & 3 & 1 & 2 & 3 & 1 & 2 & 3 & 1 & 2 & 3 \\ % inserting body of the table
\hline % inserts single horizontal line
& $\varphi_{ix}$ & 1 & 0 & 0 & 1 & 0 & 0 & 1 & 0 & 0 & 1 & 0 & 0 \\
$\Gamma_3$ & $\varphi_{iy}$ & 0 & 1 & 0 & 0 & -1 & 0 & 0 & -1 & 0 & 0 & 1 & 0 \\
& $\varphi_{iz}$ & 0 & 0 & 1 & 0 & 0 & -1 & 0 & 0 & 1 & 0 & 0 & -1 \\
\hline % inserts single horizontal line
& $\varphi_{ix}$ & 1 & 0 & 0 & -1 & 0 & 0 & -1 & 0 & 0 & 1 & 0 & 0 \\ 
$\Gamma_4$ & $\varphi_{iy}$ & 0 & 1 & 0 & 0 & 1 & 0 & 0 & 1 & 0 & 0 & 1 & 0 \\
& $\varphi_{iz}$ & 0 & 0 & 1 & 0 & 0 &  1 & 0 & 0 & -1 & 0 & 0 & -1 \\
\hline\hline %inserts single line 
\end{tabular*}
\footnotetext[1]{1-($x,y,z$), 2-($-x+\frac{1}{2},y+\frac{1}{2},z$), 3-($x+\frac{1}{2},-y+\frac{1}{2},z+\frac{1}{2}$), 4-($-x,-y,z+\frac{1}{2}$)}
\label{table:fun} % is used to refer this table in the text
\end{table}

\subsection{Model for the magnetic structure}
\begin{figure}[t]
\includegraphics[scale=1.05]{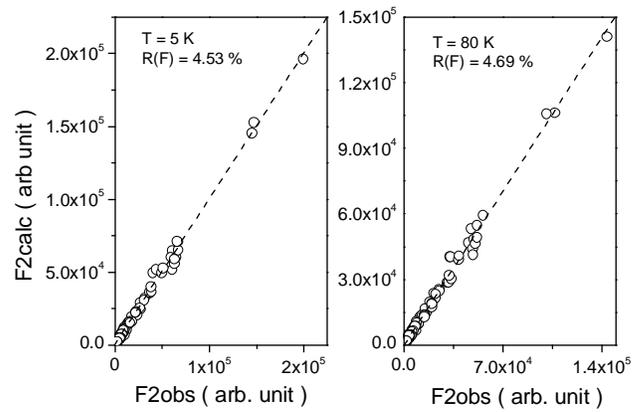}
\caption{Experimental structure factors vs. calculated ones for the refinement of the data collected at different temperatures (D10 data).}
\label{fig:fit}
\end{figure}

\begin{figure*}[t]
\includegraphics[scale=1.0]{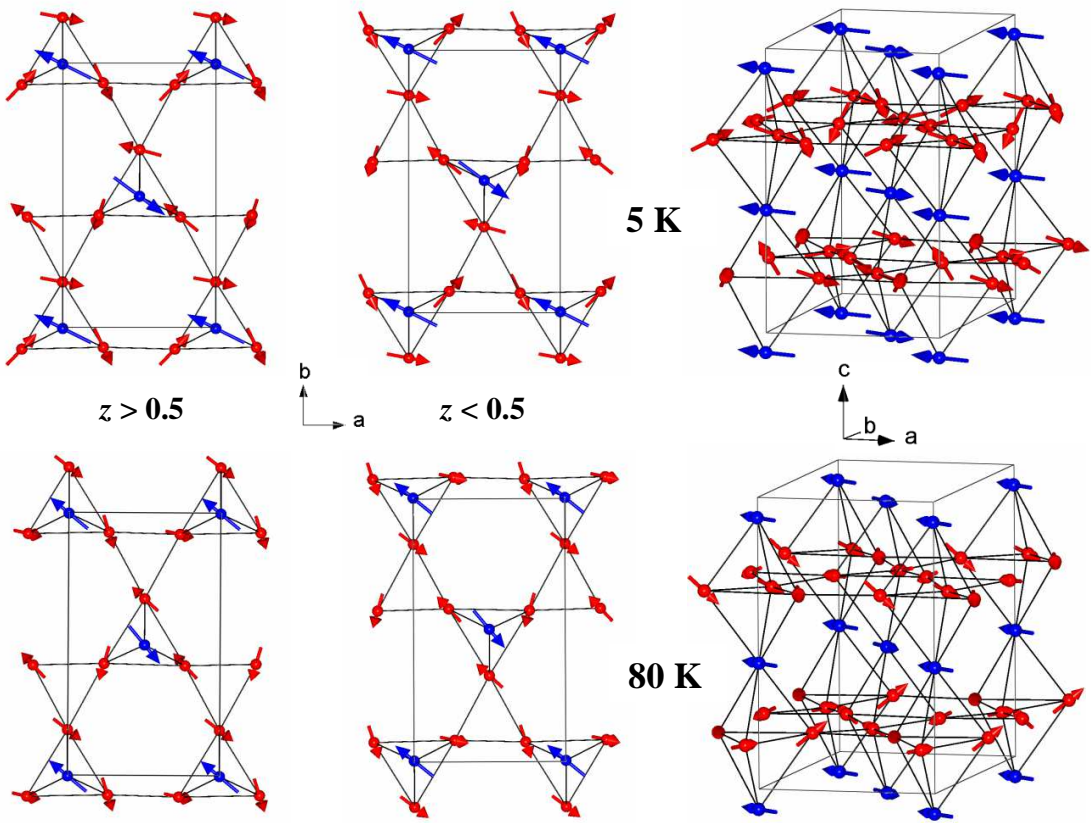}
\caption{(Color online) Schematic representation of the magnetic structure of YBaCo$_4$O$_7$ at different temperatures. The spins in triangular and Kagome sublattices are presented as large (blue) and small (red) arrows, respectively.}
\label{fig:mag}
\end{figure*}

\indent Both $\Gamma_3$ and $\Gamma_4$ irreps enters three times in the decomposition of the magnetic representation for the 4$a$ sites:
\begin{equation}
\label{eq:decomposition}
\Gamma(4a)=3\Gamma_1 \oplus 3\Gamma_2 \oplus 3\Gamma_3 \oplus 3\Gamma_4 \nonumber
\end{equation}
and therefore one needs to consider three sets of orthogonal basis vectors for each representation, that are given in Table \ref{table:fun}.\indent The admixture of the two irreps (monoclinic symmetry) splits the 4$a$ site into two independent orbits, implying that there are no constraints for magnetic moments of cobalt sites related by the $b$ and $n$ glide planes of the $Pbn2_1$ space group, and those can be treated as independent variables in the refinement. In contrary, the sites related by the $2_1'$ screw axis have parallel in-plane ($\varphi_{ix}$,$\varphi_{iy}$) configuration and an anti-parallel out-of-plane ($\varphi_{iz}$) component.\\ 
\indent Another symmetry aspect to consider is the orientation domain structure of the magnetically ordered state. Since the phase transition involves a change of Laue class, $mm2\rightarrow$2, each domain of the orthorhombic paramagnetic phase (Fig.\ref{fig:dom}) gives rise to two ferroelastic domains with lattice vectors ($\overrightarrow{a}$,$\overrightarrow{b}$,$\overrightarrow{c}$) and ($\overrightarrow{a}$,-$\overrightarrow{b}$,-$\overrightarrow{c}$), respectively. Thus, the refinement of the low temperature phase must involve six domains and eight nonequivalent Co positions. The following strategy for the magnetic structure refinement has been applied: initially, only planar spin components were allowed to vary ($z$-component was fixed to zero), in the light of results obtained previously on a polycristalline sample\cite{ISI:000242409000015}. Given the large number of variables, we have imposed the additional constraint, not derived from symmetry, that all sites in the Kagome sublattice in one hand and in the triangular sublattice on the other, have the same moment magnitudes. These simplifications, in combination with the symmetry constraints discussed above, leads to 10 independent variables. The refinement converged quickly to a spin configuration equivalent to that proposed from the powder diffraction data.\cite{ISI:000242409000015} In the second step, the $z$-components were allowed to vary, with the symmetry constraint discussed previously. The refinement included 16 magnetic parameters (98 in total) and converged with a good final agreement between calculated and measured integrated intensities (Fig.\ref{fig:fit}) and physically reasonable domain populations. The resulting magnetic structure is shown in Figure \ref{fig:mag} and the refined parameters are given in Table \ref{table:ref}. We note that an attempt to refine the magnitude of the eight spins independently was not satisfactory, most parameters becoming strongly correlated, suggesting that the best model must be obtained with the aforementioned restriction on the magnitude of the moments.\\  
\indent The magnetic model is very similar to that obtained from the powder diffraction experiment. The antiferromagnetic ordering in the triangular sublattice is practically identical in both models and no $z$-component for these spins was found from the single crystal data. However, the spins in the Kagome layers show deviations from the planar structure not initially detected in the powder measurements, with the largest angle $\sim$40$^\circ$ (Fig.\ref{fig:mag}, Table \ref{table:ref}).


\begin{figure}[b]
\includegraphics[scale=1.0]{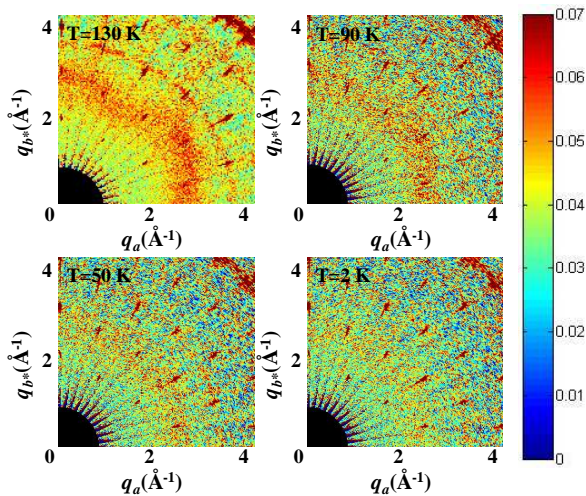}
\caption{(Color online) Neutron scattering cross sections measured in the ($ab$) plane at different temperatures (PRISMA data).}
\label{fig:diff}
\end{figure}

\begin{table*}[t]
\caption{Magnetic structure parameters refined in spherical coordinates ($M$-value of moment, $\Phi$-angle with the $x$ axis and $\Theta$-angle with the $z$ axis) at different temperatures. Coordinates for Co ions are given as they were refined in the $Pbn2_1$ space group at 150 K. Domain fractions at T=5 K and 80 K are: 0.34(1), 0.18(1), 0.11(1), 0.10(2), 0.19(2), 0.08(3) and 0.36(1), 0.18(1), 0.11(1), 0.11(2), 0.16(2), 0.08(3), respectively.}
\centering % used for centering table
\begin{tabular*}{1.00\textwidth}{@{\extracolsep{\fill}} c rr rr rr} % centered columns (4 columns)
\hline\hline %inserts double horizontal lines
Spherical components & \multicolumn{2}{c}{$M$($\mu_B$)} & \multicolumn{2}{c}{$\Phi$($^\circ$) } &  \multicolumn{2}{c}{$\Theta$($^\circ$)} \\ [0.5ex] % inserts table %heading
\hline % inserts single horizontal line
Coordinates of Co atom & 5 K & 80 K & 5 K & 80 K & 5 K & 80 K \\ % inserting body of the table
\hline % inserts single horizontal line
-0.001(4), -0.004(4), 0.931(3) & 2.62(9) & 2.04(8) & 153(9) & 140(7) & 90(-)\footnotemark[1] & 90(-)\footnotemark[1] \\
0.001(4), 0.004(4), 0.431(3) & 2.62(9) & 2.04(8) & 153(9) & 140(7) & 90(-)\footnotemark[1] & 90(-)\footnotemark[1] \\
0.501(4), 0.496(4), 0.931(3) & 2.62(9) & 2.04(8) & 153(9) & 140(7) & 90(-)\footnotemark[1] & 90(-)\footnotemark[1] \\
0.499(4), 0.504(4), 0.431(3) & 2.62(9) & 2.04(8) & 153(9) & 140(7) & 90(-)\footnotemark[1] & 90(-)\footnotemark[1] \\
-0,002(3), 0.172(3), 0.672(4) &  1.88(6) & 1.50(7) & -11(19) & 323(14) & 103(9) & 102(13) \\
0,002(3), -0.172(3), 0.172(4) &  1.88(6) & 1.50(7) & -11(19) & 323(14) & 77(9) & 78(13) \\
0.502(3), 0.672(3), 0.672(4) &  1.88(6) & 1.50(7) & -195(20) & 137(9) & 104(8) & 101(13) \\
0.498(3), 0.328(3), 0.172(4) &  1.88(6) & 1.50(7) & -195(20) & 137(9) & 76(8) & 79(13) \\
0.768(3), 0.418(2), 0.686(3) &  1.88(6) & 1.50(7) & 141(14) & 133(14) & 104(6) & 108(8) \\
0.232(3), 0.582(2), 0.186(3) &  1.88(6) & 1.50(7) & 141(14) & 133(14) & 76(6) & 72(8) \\
-0.268(3), -0.082(2), 0.686(3) &  1.88(6) & 1.50(7) & 46(18) & 348(9) & 71(8) & 128(12) \\
0.268(3), 0.082(2), 0.186(3) &  1.88(6) & 1.50(7) & 46(18) & 348(9) & 109(8) & 52(12) \\
0.264(3), 0.925(2), 0.681(3) &  1.88(6) & 1.50(7) & -65(15) & 288(14) & 116(6) & 104(5) \\
-0.264(3), -0.925(2), 0.181(3) &  1.88(6) & 1.50(7) & -65(15) & 288(14) & 64(6) & 76(5) \\
0.236(3), 0.425(2), 0.681(3) &  1.88(6) & 1.50(7) & 252(19) & 254(9) & 135(9) & 81(9) \\
0.764(3), -0.425(2), 0.181(3) &  1.88(6) & 1.50(7) & 252(19) & 254(9) & 45(9) & 99(9) \\
\hline\hline %inserts single line 
\end{tabular*}
\footnotetext[1]{variation of these parameters resulted in their small oscillations, within 2$^\circ$-3$^\circ$ interval near $\Theta$=90$^\circ$, indicating that there is no detectable $z$-component in the triangular lattice and therefore in the final refinement these parameters were fixed.}
\label{table:ref} % is used to refer this table in the text
\end{table*}
\indent From previous structural work\cite{ISI:000271351500043} it has been established that the tetrahedra in the Kagome layers make a complex tilt pattern in the orthorhombic phase. This tetrahedral tilts are not rigid modes of the structure and occur with relatively large polyhedral distortions which can essentially modify the local anisotropy of the corresponding Co sites. It appears that the spin configuration with a sizable $z$ component for certain sites is due to this local anisotropy. The average values of the moments in the triangular and Kagome lattices are 2.62(9) and 1.88(6) $\mu_B$ respectively, considerably smaller than the expected spin contribution of 3.25 $\mu_B$ for a Co ion with an average +2.25 oxidation state. This suggests that a fraction of the moment remains disordered at low temperature. Additional neutron diffraction data to probe a full 2D section of reciprocal space (Fig.\ref{fig:diff}) show that indeed diffuse scattering persists down to 2 K.\\
\indent The data collected at 80 K have been refined using the same strategy. The magnetic structure is shown in Figure \ref{fig:mag} and the refined parameters are listed in Table \ref{table:ref}. The model suggests a spin reorientation process which has been discussed previously \cite{ISI:000242409000015} based on competing in-plane and out-of-plane exchange interactions. The transformation from the 5 K structure mainly affects the Kagome sublattice and does not change the magnetic $P$112$_1'$ symmetry. Thus, the results for the oxygen stoichiometric single crystal are consistent with the powder diffraction data obtained for the same composition.\cite{ISI:000242409000015}\\
\indent  The magnetic structures refined here are obtained with less constraints than the powder data and based on many independent reflections. They allow to confirm that the model proposed in the previous single crystal study,\cite{ISI:000237969600046} where 120$^\circ$ spin configurations were assumed for both Kagome and triangular sublattices is not valid for stoechimetric samples. It is likely that the oxygen content which can vary largely in these materials is the cause of such large discrepancy between experimental results.\\
\indent Also, it is important to note that the model presented produces a small ferromagnetic moment ~0.08(6) $\mu_B$ per Co ion in the ($ab$) plane. This value is not statistically significant but the monoclinic symmetry allows for an in-plane ferromagnetic component. It is also difficult to determine whether a small ferromagnetic moment is present by magnetic measurements which are complicated by the presence of domains and the disordered (glassy) component, the latter producing non-linear magnetization that masks the saturation regime. However the results obtained here, allows us to postulate that the ferromagnetic moment found in Ca$_{1-x}$$R$$_x$Ba(Co/Fe)$_4$O$_7$ must arise from a similar non collinear spin arrangement rather than from the simple ferrimagnetic ordering with antiparallel Kagome and triangular sublattices, proposed in Refs\cite{ISI:000263733300009,ISI:000260254400003,ISI:000271744600010}. This assumption is in agreement with the recent neutron diffraction study of CaBaCo$_4$O$_7$ where a complex magnetic structure with non-collinear spin ordering in both triangular and Kagome sublattices has been found.\cite{ISI:000276207300067}\\
 \indent Finally, it should be pointed out that the magnetic point group 2$'$ allows a linear magnetoelectric coupling and is consistent with a presence of spontaneous electrical polarization along the $z$-axis. Thus, the $R$Ba(Co/Fe)$_4$O$_7$ compositions can demonstrate multiferroic/magnetoelectric properties since the necessary symmetry conditions are fulfilled.\\
\indent Thermodynamics of the phase transitions involving coupled order parameters has been discussed in a number of works.\cite{ISI:Lifshitz,ISI:A1980JK19000027,lan:dmitriev,lan:toledano} The relevant case implies two one-dimensional order parameters (noted $\eta_3$ and $\eta_4$ to follows the labeling of the irreps) with biquadratic coupling. The Landau free energy can be written: 
\begin{equation}
F(\eta_3,\eta_4)=\frac{\alpha_1}{2}\eta_3^2 + \frac{\alpha_2}{4}\eta_3^4 + \cdots + \frac{\beta _1}{2}\eta_4^2 + \frac{\beta _2}{4}\eta_4^4 + \cdots + \delta \eta_3^2 \eta_4^2 
\label{eq:FreeEnergy}
\end{equation}
where $\alpha_{1,2}$, $\beta_{1,2}$ and $\delta$ are coefficients of the expansion.
The integrity basis of $F$($\eta_3$,$\eta_4$) is formed by the two square invariants $\eta_3^2$, $\eta_4^2$ and thus, the minimal degree to 
which the expansion (1) has to be truncated to fully describe the set of symmetry distinct phases is the fourth degree.\cite{lan:dmitriev,lan:toledano} However, in this case the transition from the high symmetry phase to the phase with $\eta_3\neq$0 and $\eta_4\neq$0 can only take place in a single point of the phase diagram. A more realistic phase diagram, where these phases are separated by a first-order transition line, can be obtained by including a six-degree term in $F$($\eta_3$,$\eta_4$) for at least one of the order parameters. A comprehensive review of this case as well as the case of the symmetric six-degree expansion can be found in Refs.\cite{lan:dmitriev,lan:toledano} However, the isostructural phase transition observed in YBaCo$_4$O$_7$ does not appear either asymmetric or symmetric six-degree expansion (\ref{eq:FreeEnergy}). Nevertheless, it is straightforward to show that eight-degree asymmetric expansion can stabilize two distinct thermodynamic phases with $\eta_3\neq$0 and $\eta_4\neq$0. Minimization of (\ref{eq:FreeEnergy}) in respect of $\eta_4$ gives the equilibrium value for this order parameter:
\begin{equation}
\eta_4^2=-\frac{\beta _1+2\delta \eta_3^2}{\beta _2} 
\label{eq:minin}
\end{equation}
Substituting this expression in the $F$($\eta_3$,$\eta_4$) truncated at eight degree in respect of $\eta_3$ gives, with some renormalized polynomial coefficients, a well studied case of a single one-dimensional order parameter. For this case it has been shown \cite{lan:dmitriev,lan:toledano} that inclusion of invariants with the highest degree 2$m$ results in $m$/2 stable low-temperature phases having identical symmetry. Thus, taking into account the expression (2), the asymmetric eight-degree expansion (\ref{eq:FreeEnergy}) should split the phase space where both components of the reducible $\Gamma_3\oplus \Gamma_4$ order parameter are not zero into two distinct regions separated by the discontinuous isostructural phase transition. The eight-degree expansion necessary to account the isostructural transition observed experimentally at T$_2$ indicates that the energy difference between the two spin configurations is very small. Apparently, this situation is not unique for geometrically frustrated systems, where ground state degeneracy is lifted by a symmetry lowering. The structural distortions can favor several configurations of initially infinitely degenerated manifold making the system to be easily switchable between the different phases. This kind of systems should be sensitive for external perturbations and may demonstrate rich phase diagrams in the coordinates of thermodynamic variables such as pressure, temperature and concentration.
\section{Conclusions}
\indent An oxygen stoichiometric single crystal of YBaCo$_4$O$_7$ exhibits long-range magnetic ordering with propagation vector {\bf k}=0 below T$_1\sim$100 K. The magnetic ordering breaks the crystal symmetry $Pbn2_1$ to monoclinic symmetry $P112_1$ which has been observed by high-resolution X-ray diffraction. At T$_2\sim$60 K, another magnetic phase transition takes place, corresponding to a spin re-orientation process and with no additional symmetry breaking. In both magnetic phases (above and below T$_2$), spins in the triangular sublattice are antiferromagnetic aligned in the ($ab$) plane with no detectable z-component. On the contrary, spins in the Kagome layers are ordered in a complex non-collinear fashion and deviate from the ($ab$) plane. The transition at T$_2$ affects mainly the Kagome sublattice. The average values of spins in the triangular and Kagome layers are 2.62(9) $\mu_B$ and 1.88(6) $\mu_B$, respectively at T=5 K. The reduced values with respect to the expected spin contribution for Co in the +2.25 oxidation state is related, at least partially, to disorder and short-range correlations that persist down to 2 K, as evidenced by neutron diffuse scattering.\\
\\
\\
\section{Acknowledgment}
We would like to thank Prof. P. Toledano for helpful discussions regarding the Landau theory with coupled order parameters. Work at Argonne supported under Contract No. DE-AC02-06CH11357 by UChicago Argonne, LLC, Operator of Argonne National Laboratory, a U.S. Department of Energy Office of Science Laboratory.

\end{document}